\documentstyle[twocolumn,epsf,aps]{revtex}

\draft

\begin{document}

\title{Dynamic Scaling of Width Distribution \\
	in Edwards--Wilkinson Type Models of Interface Dynamics}

\author{Tibor Antal and Zolt\'an R\'acz}

\address{Institute for Theoretical Physics, E\"otv\"os University\\
1088 Budapest, Puskin u. 5-7, Hungary}

\address{\em (\today)}


\address{
\centering{
\medskip\em
\begin{minipage}{14cm}
{}~~~Edwards--Wilkinson type models are studied
in $1+1$ dimensions and the time-dependent distribution, $P_L(w^2,t)$,
of the square of the width of an interface, $w^2$, is calculated
for systems of size $L$. We find that, using a flat interface as an
initial condition,
$P_L(w^2,t)$ can be calculated exactly and it obeys
scaling in the form $\langle w^2 \rangle_\infty P_L(w^2,t) =
\Phi ({w^2 / \langle w^2 \rangle_\infty} ,{t / L^2}) $
where $\langle w^2 \rangle_\infty$ is the stationary value of $w^2$.
For more complicated initial states, scaling is observed only
in the large-time limit and the scaling function depends on the
initial amplitude of the longest wavelength mode.
The short-time limit is also interesting since $P_L(w^2,t)$
is found to closely approximate the log-normal distribution.
These results are confirmed by Monte Carlo simulations on a
`roof-top' model of surface evolution.
\end{minipage}
}
}

\maketitle

\narrowtext

\section{Introduction}

Interfaces play an important role in a number of physical,
chemical, and biological phenomena.
Their fluctuations display universal features and, accordingly, models
of interface motion have been analyzed and distinguished in terms
of universality classes \cite{Krug}.
The classification usually proceeds by
taking systems of various sizes $L$ and measuring the time-evolution
of the average of the square width, $\langle w^2 \rangle$, of the interface.
Then the values of the static $(\zeta)$ and dynamic $(z)$ exponents
which determine the universality class are obtained \cite{Fam}
by observing collapse of data in accordance with the
scaling form $\langle w^2 \rangle \sim L^{2 \zeta} f(t/L^z)$.

In practice, this procedure is not so easy to realize since
$\langle w^2 \rangle$ is an integral over all modes in
the system and, consequently, large corrections to scaling are present.
Recently, it was suggested \cite{foltin:1994} that an alternative and
more detailed characterization of interfaces may be obtained
through the probability distribution
of the random variable $w^2$.
The steady-state distribution function $P_L^{(s)}(w^2)$ has been
calculated exactly for various growth models
\cite{{foltin:1994},{PRZ:1994},{PR:1994}} and it has been found that
$P_L^{(s)}(w^2)$ defines a scaling function, $\Phi^{(s)}$,

  \begin{equation}
  \langle w^2 \rangle_\infty P_L^{(s)}(w^2) = \Phi^{(s)}
  (w^2/\langle w^2 \rangle_\infty)
  \quad
  \label{statscal}
  \end{equation}
which is a universal characteristic of the interface fluctuations.

The above scaling function, however, distinguishes only among
static universality classes. For example, $\Phi^{(s)}(x)$
is the same for both the
one-dimensional ($d=1$) Edwards-Wilkinson (EW) \cite{ew:1982} and
Kardar-Parisi-Zhang (KPZ) \cite{kpz:1986} equations since these
models describe processes which have the same steady-state distributions and
differ only in the scaling of their dynamics.
Thus, in order to distinguish among dynamical universality classes
one should extend the static results to the
time-dependent width distributions, $P_L(w^2,t)$. This is what we shall do here
for the simplest model of surface growth, the $d=1$ EW
equation \cite{ew:1982}.

Since the EW equation is linear, much of the calculation can be done
analytically (Section \ref{analytical}) and, in particular,
we can derive $P_L(w^2,t)$ and the associated dynamical scaling function
$\langle w^2 \rangle_\infty P_L(w^2,t)$ in closed form
for the case of a flat initial surface. An interesting feature of the
result is that the short-time limit is closely approximated by the
log-normal distribution (Section \ref{short-long}).
Arbitrary initial conditions are harder to treat and we obtain general results
only for the long-time limit where the dependence on the
initial-state disappears except for the initial
amplitude of the longest wavelength mode (Section \ref{short-long}).

In order to carry out a limited check of the universality of our
results we used Monte Carlo simulations to study a
`roof-top' model of surface evolution
\cite{{Meakin:1986},{prl:1986}}.  This model belongs to the EW universality
class when the overall velocity of the surface is zero. Otherwise,
it is in one universality class with the KPZ equation.
Excellent agreement is found (Section \ref{MC}) between the analytic and
simulation results for the dynamical scaling function in the EW limit
and, furthermore, we also find that this scaling function is
easily distinguishable from the corresponding function
obtained for the KPZ case.

\section{Calculation of the width distribution}
\label{analytical}

A simple model of surface evolution governed by surface tension and noise
is the EW equation \cite{ew:1982}:
  \begin{equation}
  {\partial h(x,t) \over \partial t} = \nu
  {\partial^2 \over \partial x^2} h(x,t) + \eta(x,t) \ .
  \label{eq:ew}
  \end{equation}
Here  $h(x,t)$ is the height of the surface
at sites $0\le x\le L$,
$\nu$ is a constant related to the dynamical surface tension,
and $\eta$ is a Gaussian white noise of strength $\Gamma$:
  \begin{equation}
  \langle \eta (x,t) \eta (x',t') \rangle
  = 2 \Gamma \delta (x-x') \delta (t-t') \ .
  \label{eq:zaj}
  \end{equation}
For simplicity, we shall assume periodic boundary conditions.
It should be noted, however,
that free boundary conditions may be more realistic in higher
dimensions where comparisons with experiments are possible.

Our aim is to calculate the time-dependent distribution, $P_L(w^2,t)$,
for an arbitrary initial condition, $h(x,0)$.
The derivation follows along the line that has been worked out for the
static case
\cite{foltin:1994} with extra complications arising from time dependence
as well as from initial conditions.

The quantity $w^2(t)$ is defined for
a configuration $h(x,t)$ as the mean square fluctuations
of the height:

  \begin{equation}
  w^2(t) = \overline{h^{\, 2}} - \overline{h}^{\, 2}\quad ,
  \label{eq:w2def}
  \end{equation}
where the time-dependent average, ${\overline f}(t)$, of a function $f(x,t)$
is obtained as its spatial average

  \begin{equation}
  \overline{f}(t)=\frac{1}{L} \int_{0}^{L} dx \,f(x,t) \quad .
  \label{eq:hbar}
  \end{equation}
The first step of the calculation of $P_L(w^2,t)$ is writing it in terms of a
path integral
  \begin{equation}
   P_L(w^2,t)=\int {\cal D}[h]\
   \delta [w^2 - (\overline{h^2} - \overline{h}^2) ]
   \ p(\{ h\},t)
   \label{pathint}
   \end{equation}
where $p(\{ h\},t)$ is the path-probability that
the surface evolves from an initial state $h(x,0)$ to a
configuration $h(x,t)$ in time $t$. The dependence on the
initial condition is not written explicitly though it is understood
that this dependence is important for any finite $t$.

The Laplace transform of equation (\ref{pathint}) gives the
generating function of the moments of $P_L(w^2,t)$:
\begin{equation}
G_L(\lambda ,t)  = \int_{0}^{\infty} d\zeta \,P_L(\zeta , t )
e^{-\lambda \zeta} \quad ,
\end{equation}
and one finds that $G(\lambda ,t )$ is the following path integral
\begin{equation}
G_L(\lambda ,t) =  \int {\cal D}[h] \ p(\{ h\},t)
\exp\left[ - \lambda \,(\,
\overline{h^{\, 2}} \,-\, \overline{h}^{\, 2} \, ) \right] \ .
\label{Glambda}
\end{equation}

The next step is to note that the above path integral can be
written as an infinite product of ordinary integrals provided the
system is described in terms of Fourier amplitudes. Indeed, let us
write
  \begin{equation}
  h(x,t)-\overline h(t) = \sum_{n=-\infty}^\infty c_n(t) e^{ i k_n x }\ ,
  \end{equation}
where $k_n=2\pi n /L$ and $c_{-n} = c_n^*$ (note that $c_0\equiv 0$, thus
the $n=0$ Fourier mode can be left out from further considerations).
The EW equation is replaced now by an infinite set of ordinary differential
equations:
  \begin{equation}
  \dot c_n(t) + \nu k_n^2 c_n(t) = \eta_n(t) \ ,
  \label{diff}
  \end{equation}
where $\eta_n(t)$, defined as the
Fourier transform of $\eta (x,t)$, is also an uncorrelated white noise.
We can see that the Fourier modes with different $n$-s are decoupled
and evolve independently. Thus the probability of a path that an
initial state characterized by a set of $\{ c_n(0)\}$ evolve into
$\{ c_n(t)\}$ is just the product of probabilities
$p_n[c_n(t)|c_n(0)]$ that $c_n(0)$ evolves into $c_n(t)$:
\begin{equation}
\hat p[\{ c_n(t)\}|\{ c_n(0)\}]=\prod_{n=-\infty}^\infty p_n[c_n(t)|c_n(0)]
\quad .
\label{prod}
\end{equation}
Equation (\ref{diff}) is well known as the Langevin equation of Brownian
motion and thus $p_n[c_n(t)|c_n(0)]$ can be obtained from the Fokker-Planck
description of the process \cite{Healy:1994}:
  \begin{equation}
  p_n[c_n(t)|c_n(0)]= { 1 \over 2\pi \sigma_n^2(t) }
  \exp \left[-{ |c_n - \langle c_n(t) \rangle |^2 \over
  2\sigma_n^2(t) }\right] \ ,
  \label{pcn}
  \end{equation}
where:
  \begin{equation}
  \langle c_n(t) \rangle = c_n(0) e^{-\nu k_n^2 t}\ , \
  \sigma_n^2(t) = {\Gamma ( 1 - e^{-2 \nu k_n^2 t} )\over L \nu k_n^2}  \ .
  \label{eq:cn-sign}
  \end{equation}

Now we can write the functional integral (\ref{Glambda}) as a product
of integrals over the coefficients $c_n$:
  \begin{equation}
  G_L(\lambda,t) = \prod_{n=1}^\infty \int dc_n dc_n^*
  p_n^2\left[c_n(t)|c_n(0)\right] e^{-2\lambda |c_n|^2} \ .
  \end{equation}
Substituting $p[c_n(t)|c_n(0)]$ from (\ref{pcn}), the Gaussian
integrals can be calculated and the inverse Laplace transform of the result
gives the
time-dependent probability distribution in a scaled form which does not
depend explicitely on $L$:
  $$
  \langle w^2 \rangle_\infty P_L(w^2,t)
  \equiv \hat\Phi (x,\tau ,\{ s_{n0}\} ) = \nonumber
  $$
  \begin{equation}
  \int\limits_{-i\infty}^{i\infty}{ dy \over 2 \pi i}
   \ e^{xy} \prod_{n=1}^\infty
  { \exp\left[-y \ s_{n0}^2 \ e^{-n^2\tau}/(1 + ya_n) \right ]
	  \over
  (1 + y a_n) }	\ .
  \label{eq:intform}
  \end{equation}
Here $\langle w^2\rangle_\infty =L\Gamma/(12\nu)$ and
  \begin{equation}
  a_n = {6 \over (\pi n)^2} (1-e^{-\tau n^2}) \ ,
  \end{equation}
with the scaling variables given by
  \begin{equation}
  x= {w^2 \over \langle w^2 \rangle_\infty}\ , \
  \tau= {8 \pi^2 \nu t \over L^2}\ , \
  s_{n 0}^2={ 2|c_n (0)|^2 \over \langle w^2 \rangle_\infty } \ .
  \label{eq:xtau}
  \end{equation}

\begin{figure}[htb]
\centerline{
	\epsfxsize=9.5cm
	\epsfbox{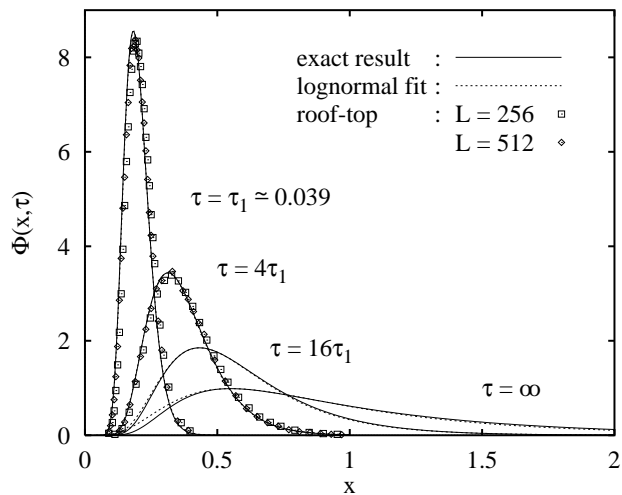}
		\vspace*{0.5cm}
	   }
\caption{Evolution of the scaling function of width distribution
         in case of a flat initial interface [eq.(18)].
         The scaling variables $x$ and $\tau$ are given
         in eq.(17). Analytical results for the EW model are compared
	 with log-normal fits
	 and with Monte Carlo results on the `roof-top' model.
	 The exact results and the lognormal fits are indistinguishable within
         linewidth for $\tau\le 4\tau_1$.}
\label{fig:phi}
\end{figure}

In case of flat initial surface $(s_{n0}=0)$, one can evaluate the
integral (\ref{eq:intform})
exactly by collecting contributions from simple poles at $-1/a_n$ and
one finds a scaling function of two variables:
  \begin{eqnarray}
  &\Phi(x,\tau)\equiv& \hat\Phi(x,\tau ,\{0\})=
      \nonumber \\
  &&\sum\limits_{m=1}^\infty {1 \over a_m}
  \exp \left\{ {- x \over a_m} \right\}
  \prod\limits_{n=1, n\ne m}^\infty { a_m \over a_m - a_n} \
  \label{eq:phiexact}.
  \end{eqnarray}
For any finite $x$, the above sum can be
approximated by a finite number of terms and $\Phi(x,\tau)$
can be evaluated with a given accuracy.
Difficulties arise only in the limit of $\tau\rightarrow 0$
where $\Phi(x,\tau)$ goes into a delta-function.
The evolution of $\Phi(x,\tau)$ obtained by numerical evaluation
of (\ref{eq:phiexact}) can be seen in Fig.~\ref{fig:phi}.

We conclude this section by calculating
the time dependence of $\langle w^2 \rangle$ which will be
needed for comparing the time-scale of MC simulations with the time-scale
of the EW equation. We find $\langle w^2 \rangle_t$
from the generating function as
$ - \partial_\lambda G(\lambda) |_{\lambda=0} $ \ :
  \begin{equation}
  { \langle w^2 \rangle_t \over \langle w^2 \rangle_\infty }
  = 1 + \sum_{n=1}^\infty \left(s_{n0}^2-{6 \over \pi^2n^2}\right)
  e^{-\tau n^2} \ .
  \label{eq:w2t}
  \end{equation}
In the long time limit $(e^{-\tau} \ll 1)$
this expression reduces to
  \begin{equation}
  { \langle w^2 \rangle_t \over \langle w^2 \rangle_\infty } =
  1 + ( s_{10}^2 - {6 \over \pi^2} ) e^{-\tau}
  + {\cal O}( e^{-4\tau}) \ ,
  \label{eq:w2longt}
  \end{equation}
and one can see that $\langle w^2 \rangle_t$ approaches its steady-state
value much faster if we start
with an initial surface where $s_{10}^2$ is set to its steady-state value
$s_{10}^2=s_{1\infty}^2= 6/\pi^2$. The above equation
also suggest a method for finding out in simulations if the system
has settled to its steady-state. One can choose small
($s_{10}^2\ll s_{1\infty}^2$) and large
($s_{10}^2\gg s_{1\infty}^2$) initial values for $s_{10}^2$ and then
$\langle w^2 \rangle_t$ approaches its steady-state value from below and above,
correspondingly.
If the two values converged, one may assume that the
steady state has been reached. This type of checking
for steady state is widely used in simulations of
equilibrium systems such as the Ising model where completely ordered
and disordered initial states are employed. Similar
procedures, however, do not seem to have been followed in the simulations of
surface evolution models.

For short times, one can change the sum (\ref{eq:w2t}) into an integral and
$\langle w^2 \rangle_t - \langle w^2 \rangle_{t=0}$ can also be calculated
  \begin{equation}
  \langle w^2 \rangle_t =
  \langle w^2 \rangle_0 +
  \sqrt{ { 2 \over \pi \nu } } \Gamma t^{1/2}
  + {\cal O}(t) \ .
  \end{equation}
The above result is valid for a flat initial state as well as for
an initial surface containing
finite number of nonzero Fourier terms ($s_{n0}\not= 0$ for $n\leq n_{max}$).

\section{Long- and short-time asymptotics}
\label{short-long}

The case of arbitrary initial conditions with nonzero $s_{n0}$-s
is complicated by the presence of
essential singularities in the function which is integrated in
(\ref{eq:intform}). As a consequence, we are not able
to evaluate the probability distribution in general.
The representation (\ref{eq:intform}) is useful, however for finding the
long- and short-time asymptotics of $\Phi(x,\tau)$.

For $\tau\rightarrow \infty$, one has to keep contributions which are
proportional to $e^{-\tau}\ll 1$. It follows then that $\Phi$ depends
on the initial state only through the initial amplitude, $s_{10}$, of the
longest wavelength mode. The calculation of the terms which are
proportional to $e^{-\tau}$ involves collecting
contributions from both
simple poles and quadratic singularities in (\ref{eq:intform})
with the result that the $s_{10}$ dependence appears in a prefactor in front
of a new scaling function $\Psi(x)$:
  \begin{eqnarray}
  &\hat\Phi(x,\tau &,\{s_{n0}\}) = \nonumber \\
  &&\Phi_s(x) +
  \left(1 - {s_{10}^2 \pi^2 \over 6}\right)
  \Psi(x) \ e^{-\tau} +{\cal O}( e^{-2\tau}) \ ,
  \label{Phi}
  \end{eqnarray}
where
  \begin{eqnarray}
  \Psi(x) &=& {\pi^2 \over 3} \left( {7 \over 4} - {\pi^2 \over 6} x \right)
  \exp \left\{ - { \pi^2 \over 6 } x \right\}
				\nonumber  \\
  &-& { \pi^2 \over 3 } \sum_{m=2}^\infty (-1)^{m-1}
  { m^4 \over 1 - m^2 }
  \exp \left\{ - {\pi^2 \over 6} m^2 x\right\} \ .
  \label{Psi}
  \end{eqnarray}
This scaling function $\Psi(x)$ is shown in Fig.~\ref{fig:psi} and
discussed in Section \ref{MC}.

Comparing equation  (\ref{Phi}) with (\ref{eq:w2longt}), one can see
that the same prefactor, $s_{10}^2 - 6 /\pi^2$, appears in front of
$e^{-\tau}$ in both cases. Thus the relaxation of the scaling function
is also accelerated
if $s_{10}$ is chosen to be the steady-state value $s_{1\infty}$. Furthermore,
it also follows that the steady-state distribution can also be bracketed
by choosing small and large initial values for $s_{10}$.

Now we turn to the description of the short-time limit
of $\Phi(x,\tau)$ in case of a flat initial condition.
The description is based on an earlier observation \cite{RW} that
the fluctuations of chemical reaction fronts which are supposed to
belong to the EW universality class produce a $\Phi(x,\tau)$ which
is rather well approximated by log-normal distribution \cite{lognorm}
  \begin{equation}
  \Phi (x,\tau) \approx {\cal L}(x, x_0, \sigma )=
  {1 \over \sqrt{2 \pi} \sigma x }
  \exp\biggr\{ {-{\ln^2(x/x_0) \over 2 \sigma^2 } } \biggr\} \quad ,
  \end{equation}
where $x_0(\tau )$ and $\sigma (\tau )$ are fitting parameters which can be
determined from various considerations. We have
determined $x_0(\tau)$ and $\sigma(\tau)$ by equating both the maxima and
positions of the maxima $x_m$ of the two functions
$\Phi(x,\tau)$ and ${\cal L}(x, x_0, \sigma )$.
The values of $x_m$ and $\Phi(x_m,\tau)=\Phi_m(\tau )$ have been determined
numerically from (\ref{eq:phiexact}) and then $\sigma$
was obtained by solving the following equation:
  \begin{equation}
  \sqrt{2 \pi} x_m\Phi_m \sigma =e^{-\sigma^2 /2} \quad ,
  \end{equation}
and finally, $x_0$ was expressed as $x_o=x_m \exp{(\sigma^2)}$.
For sufficiently short
times, the width of the distribution goes to zero and
$\sigma \rightarrow 0$. In this limit the
expressions for $x_0$ and $\sigma$ simplify to
$x_0 \approx x_m$, and $\sigma \approx 1/(\sqrt{2 \pi} x_m \Phi_m)$.

\begin{figure}[htb]
\centerline{
	\epsfxsize=9.5cm
	\epsfbox{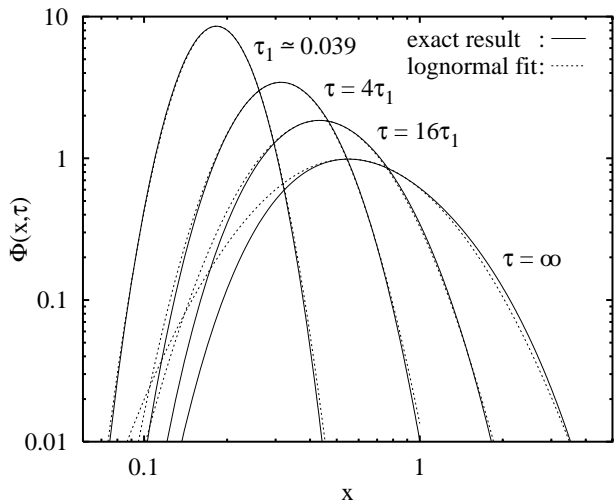}
		\vspace*{0.5cm}
	   }
\caption{ The analytical results
	and the log-normal fits of Fig.1 on a log-log plot.
	}
\label{fig:philog}
\end{figure}

The results of this fitting procedure can be seen on Fig.~\ref{fig:phi}.
Qualitatively, the fit is quite good over the whole time interval and
it clearly becomes excellent at short times. In order to make the
quality of the short-time fit more apparent, we have redrawn the curves
of Fig.~\ref{fig:phi} on a log-log plot with the results shown on
Fig.~\ref{fig:philog}. As one can see again, the log-normal fit becomes better
at small times and, at $\tau =0.039$, the fit becomes
practically indistinguishable from
$\Phi (x,\tau)$ in an interval where the function decreases
from its maximum value by three orders of magnitude.

A more quantitative description of the quality
of the log-normal fit can be given by defining a relative distance
between the two functions as
\begin{equation}
\ell (\tau) =\max_x
\frac{|\Phi (x,\tau) - {\cal L}(x, x_0, \sigma )|}{\Phi_m(\tau)} \quad .
\end{equation}
This distance increases with $\tau$ and reaches its maximum value
$\ell_{max}=\ell (\infty )\approx 0.14$ in the stationary state.
For $\tau\rightarrow 0$, we find
that $\ell \sim\tau^{1/2}$ and $\ell <0.01$ for $\tau<0.02$.
Although the diminishing relative distance actually comes from the ratio of a
strongly divergent $\Phi_m(\tau)\sim\tau^{-3/4}$
and a less divergent maximum
distance $\max_x |\Phi - {\cal L}|\sim \tau^{-1/4}$,
plots of the two functions
which extend from zero to their maxima are indistinguishable
(see Fig.\ref{fig:phi}) for $\tau <0.1$.

\section{Monte Carlo simulations}
\label{MC}

In order to see if the dynamic scaling found for
the width distribution had the expected universality, we carried out
Monte Carlo simulations for a `roof-top' model of surface
evolution \cite{{Meakin:1986},{prl:1986}}.
In this model the height of the surface
is characterized by a single-valued function $h_i$ at sites
$i=1,2,...,L$ and periodic boundary conditions $h_{i+L}=h_i$
are imposed. The height differences are restricted to
$h_{i+1}-h_i = \pm 1$ and the evolution consists of particles
being deposited at local minima or evaporating from local maxima
of the surface with rates $p_+$ and $p_-=1-p_+$, respectively.
If $p_+=p_-=1/2$,
the model belongs to the universality class of EW model while
for $p_+\not= p_-$ the universality class is that of the
KPZ equation \cite{prl:1986}.

For equal rates, one can obtain \cite{prl:1986} an exact expression for
$\langle w^2 \rangle_t$ and comparing the result
with the solution of the EW equation (\ref{eq:w2t}),
the time-scale of the MC simulation can be related to that of the EW equation.
In this way one finds that the parameters $\nu$ and $\Gamma$
should be set to $\nu =\Gamma =1/2$.
Then the $x$-s and $\tau$-s in the MC data and in the EW equation are
related in a unique way and there is no parameters to fit
when the $\Phi$-s are compared. Fig.~\ref{fig:phi}
shows both the $\Phi(x,\tau)$-s obtained from simulation and
the theoretical curves of the EW model. One finds good agreement although
a small systematic shift of the MC curves towards larger values of
$x$ can be observed. This shift is due to the
fact that, in the `roof-top' model, the initial surface is
not entirely flat ($w^2_{t=0}=1/4$)
in variance with the $w^2_{t=0}=0$ used in the theoretical calculation.
This difference should disappear
in the $L\rightarrow \infty$ limit and, indeed, one can see that the
difference is smaller for the $L=512$ sample as compared to the $L=256$
system.

\begin{figure}[htb]
\centerline{
	\epsfxsize=9.5cm
	\epsfbox{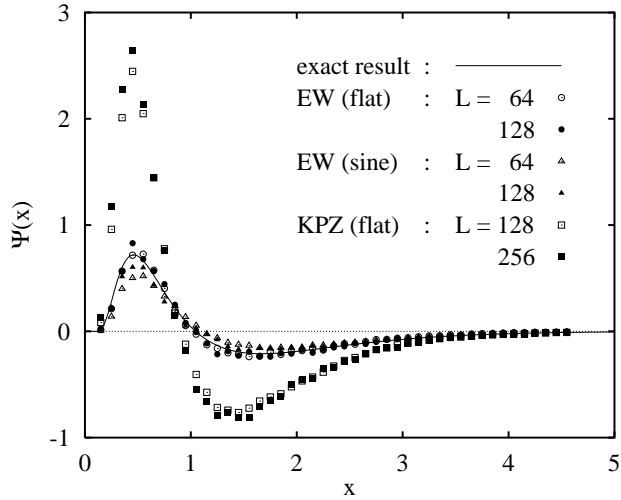}
		\vspace*{0.5cm}
		   }
\caption{ Scaling function, $\Psi(x)$, describing the long-time
        relaxation of the width distribution [eq.(22)].
        The theoretical curve for the EW model is compared with
	Monte Carlo results for both the EW and KPZ limits
	of the `roof-top' model.
	The initial states are either flat or contain a single
	sine perturbation with the longest available wavelength.
	}
\label{fig:psi}
\end{figure}

We have also examined the function $\Psi(x)$ which characterizes the
scaling of the long time
relaxation of the distribution function in the EW model (\ref{Psi}).
Since one can find a large enough time-window where
$\langle w^2 \rangle_\tau - \langle w^2 \rangle_\infty \sim
e^{-\tau}$ for all x, the function $\Psi(x)$
can be determined accurately and,
as can be seen from Fig.~\ref{fig:psi}, there is an excellent agreement
with the theoretical curve.
Thus we can conclude that the `roof-top' model which belongs to the
EW universality class for $p_+=p_-=1/2$
indeed produces the same
time dependent distribution $\Phi(x,\tau)$ as the EW equation.

In order to investigate if the $\Phi(x, \tau )$-s characterizing the
EW and the KPZ classes were distinguishable, we
have also studied the long time behavior of the `roof-top' model
for unequal rates ($p_+=1$, $p_-=0$).
In this case, we find that
$\langle w^2 \rangle_\infty P_L(x,t) - \Phi(x,\infty)$
decays with time exponentially, $\exp(-\alpha_L t)$, with a
relaxational rate, $\alpha_L$ independent of x
and, as can be seen on Fig.~\ref{fig:psi},
the coefficient, $\Psi(x)$, of the exponential differs
significantly from that of the corresponding EW scaling function.

\section{Final remarks}

It has been demonstrated previously \cite{{foltin:1994},{PRZ:1994},{PR:1994}}
that one can build a `picture gallery' of scaling functions for
steady-state width distributions and this gallery may be used for
distinguishing the static universality classes of growth processes. Here we
have made the first steps towards building a similar gallery
for dynamic scaling functions and we believe that this gallery will be
equally instrumental in recognizing dynamical universality classes.
At this moment we have results only for the one-dimensional EW and
KPZ processes but there does not seem to be any principal difficulty
in extending these calculations to other processes and to
higher dimensions by using exact solutions, renormalization-group methods,
and simulations.

An interesting byproduct of our calculation is the result that
the early-time width distribution in the EW process is practically
identical to the lognormal distribution. Lognormal-like distributions
tend to emerge more often in biological and social sciences than in
physics \cite{{lognorm},{Montroll}} and they are usually understood in terms
of the `law of proportionate effect' or on the basis of the assumption
that an event occurs only if a large number of independent `sub-events' take
place.
In our case, the lognormal distribution is produced by EW dynamics
and it appears as a characteristic of the initial
roughening of an interface.
Whether this generation of lognormal-like  distributions was
new or it was equivalent to one of the standard derivations
remains to be understood.

\section*{Acknowledgments}
We thank M. Plischke and R.K.P. Zia for helpful conversations.
This work was supported by the
Hungarian Academy of Sciences Grant N$^o$ OTKA 2090,
and by EC Network Grant N$^o$ ERB CHRX-CT92-0063.

\end{document}